\def\papertitle{One Billion Audio Sounds \\
from GPU-enabled Modular Synthesis}
\def\paperauthorA{Joseph Turian}
\def\paperauthorB{Jordie Shier}
\def\paperauthorC{George Tzanetakis}
\def\paperauthorD{Kirk McNally}
\def\paperauthorE{Max Henry}

\documentclass[twoside,a4paper]{article}
\usepackage{etoolbox}
\usepackage{dafx_20}
\usepackage{amsmath,amssymb,amsfonts,amsthm}
\usepackage{euscript}
\usepackage[T1]{fontenc}
\usepackage[utf8]{inputenc}
\usepackage{nimbusserif}
\usepackage{ifpdf}
\usepackage[english]{babel}
\usepackage{caption}
\usepackage{subfigure} %
\usepackage{amssymb}
\usepackage{color}
\usepackage{tabularx}
\usepackage{array}

\ifdefined\showmargins 
\usepackage{showframe}
\newlength\Fcolumnseprule
\makeatletter
\newcommand\ShowInterColumnFrame{
\patchcmd{\@outputdblcol}
 {{\normalcolor\vrule \@width\columnseprule}}
 {\vrule \@width\Fcolumnseprule\hfil
   {\normalcolor\vrule \@width\columnseprule}
   \hfill\vrule \@width\Fcolumnseprule
 }
 {}
 {}
}
\makeatother
\ShowInterColumnFrame
\fi

\input glyphtounicode
\ninept

\newcounter{numauth}
\newcounter{listcnt}
\newcommand\authcnt[1]{\ifdefined#1 \stepcounter{numauth} \fi}

\newcommand\addauth[1]{
\ifdefined#1 
\stepcounter{listcnt}
\ifnum \value{listcnt}<\value{numauth}
\appto\authorslist{, #1}
\else
\appto\authorslist{~and~#1}
\fi
\fi}
\authcnt{\paperauthorB}
\authcnt{\paperauthorC}
\authcnt{\paperauthorD}
\authcnt{\paperauthorE}
\authcnt{\paperauthorF}
\authcnt{\paperauthorG}
\authcnt{\paperauthorH}
\authcnt{\paperauthorI}
\authcnt{\paperauthorJ}
\def\authorslist{\paperauthorA}
\addauth{\paperauthorB}
\addauth{\paperauthorC}
\addauth{\paperauthorD}
\addauth{\paperauthorE}
\addauth{\paperauthorF}
\addauth{\paperauthorG}
\addauth{\paperauthorH}
\addauth{\paperauthorI}
\addauth{\paperauthorJ}

\usepackage{times}

\newif\ifpdf
\ifx\pdfoutput\relax
\else
   \ifcase\pdfoutput
      \pdffalse
   \else
      \pdftrue
\fi

\ifpdf %
  \usepackage[pdftex,
    pdftitle={\papertitle},
    pdfauthor={\authorslist},
    pdfsubject={Proceedings of the 23rd International Conference on Digital Audio Effects (DAFx-20)},
    colorlinks=false, %
    bookmarksnumbered, %
    pdfstartview=XYZ %
  ]{hyperref}
  \usepackage[pdftex]{graphicx}
\else %
  \usepackage[dvips]{epsfig,graphicx}
  \usepackage[dvips,
    pdftitle={\papertitle},
    pdfauthor={\authorslist},
    pdfsubject={Proceedings of the 23rd International Conference on Digital Audio Effects (DAFx-20)},
    colorlinks=false, %
    bookmarksnumbered, %
    pdfstartview=XYZ %
  ]{hyperref}
\fi
\usepackage[hypcap=true]{caption}
\title{\papertitle}

\threeaffiliations{
\, \,\paperauthorA\,\sthanks{Equal contribution}}
{{Spooky Audio}\\ ~\\ Berlin, Germany\\
{\tt \href{mailto:lastname@gmail.com}{lastname@gmail.com}}
}
{\paperauthorB\ \footnotemark[1], \paperauthorC\,, \paperauthorD\ }
{\href{https://www.uvic.ca/finearts/music/graduate/our-programs/music-csc/}{Computer Science and Music Technology} \\ University of Victoria\\ Victoria, Canada\\
{\tt \href{mailto:jordieshier@gmail.com}{jordieshier@gmail.com}}
}
{\paperauthorE \,}
{\href{https://mt.music.mcgill.ca/}{Music Technology Area} \\ McGill University \\ Montreal, Canada \\ {\tt \href{mailto:max.henry@mail.mcgill.ca}{max.henry@mail.mcgill.ca}}
}

\begin{document}
\ifpdf %
  \DeclareGraphicsExtensions{.png,.jpg,.pdf}
\else  %
  \DeclareGraphicsExtensions{.eps}
\fi

\maketitle

\begin{abstract}
We release synth1B1, a multi-modal audio corpus consisting of 1 billion 4-second synthesized sounds, paired with the synthesis parameters used to generate them. The dataset is 100x larger than any audio dataset in the literature. We also introduce torchsynth, an open source modular synthesizer that generates the synth1B1 samples on-the-fly at 16200x faster than real-time (714MHz) on a single GPU.
Finally, we release two new audio datasets: FM synth timbre
and subtractive synth pitch.
Using these datasets, we demonstrate new rank-based evaluation criteria for existing audio representations. Finally, we propose a novel approach to synthesizer hyperparameter optimization.

\end{abstract}

\vspace{1mm}
\section{Introduction}
\label{sec:intro}

\begin{sloppypar}
Machine learning (ML) progress has been driven by training regimes that leverage large corpora. The past decade has seen great progress in natural language processing (NLP) and vision tasks using large-scale training. As early as 2007, Google \cite{brants-etal-2007-large} achieved state-of-the-art machine translation results using simple trillion-token $n$-gram language models.
Recent work like GPT3 \cite{NEURIPS2020_1457c0d6} suggests that it is preferable to do less than one epoch of training on a large corpus, rather than multiple epochs over the same examples. Even tasks with little training data can be attacked using self-supervised training on a larger, related corpus followed by a transfer-learning task-specific fine-tuning step.
\end{sloppypar}

\begin{table}[thb]
\begin{center}
\begin{tabular}{r|c|r|c}
Name & Type & \#hours & Multi-modal \\
\hline
Diva \cite{esling2020flow} & synth & 12 & parameters \\
FSDK50 \cite{fonseca2020fsd50k} & broad & 108 & tags \\
NSynth \cite{engel2017neural} & notes & 333 & tags \\
Amp-Space \cite{naradowsky2020amp} & guitar fx & 525 & parameters \\
LibriSpeech \cite{librispeech} & speech & 1000 & text \\ 
DAMP-VPB \cite{smule_inc_2017_2616690} & songs & 1796 & lyrics \\
Audioset \cite{45857} & broad & 4971 & video+tags \\ 
YFCC100M \cite{thomee2016yfcc100m} & broad & 8081 & video \\
Libri-Light \cite{librilight} & speech & 60000 & weak labels \\
MSD \cite{bertin2011million} & songs & 72222 & tags+metadata \\
Jukebox \cite{dhariwal2020jukebox} & songs & 86667 & {\footnotesize lyrics+metadata}\\
synth1B1 & synth & 1111111 & parameters \\
\end{tabular}
\end{center}
\caption{Large-scale and/or synthesizer audio corpora.}
\label{tbl:audio-corpora}
\end{table}

Audio ML research tends to lack many large-scale corpora, instead involving multiple epoch training on comparably small corpora compared to vision or NLP. Table~\ref{tbl:audio-corpora} summarizes various large-scale and/or synthesizer audio corpora.
Using AudioSet \cite{45857} requires scraping 5000 hours of Youtube videos, many of which become unavailable over time (thus impeding experimental control). FSD50K \cite{fonseca2020fsd50k}, a free corpus, was recently released to mitigate these issues, but contains only 108 hours of audio. To our best knowledge, the largest audio set used in published research is Jukebox \cite{dhariwal2020jukebox}, which scraped 1.2M songs and their corresponding lyrics. Assuming average song length is 4:20, we estimate their corpus is $\approx$90K hours.

In this paper we introduce synth1B1, a multi-modal audio corpus consisting of 1 billion 4-second synthesized sounds, which is 100x larger than any audio dataset in the literature. The dataset is multi-modal in that each sound is paired with the corresponding synthesizer parameters used to generate it. To create such a large dataset, we built torchsynth,\footnote{\url{https://github.com/torchsynth/torchsynth}} a GPU-based open source modular synthesizer that builds the synth1B1 samples on-the-fly at 16200x faster than real-time. 
In order to compare the scope of sounds covered by synth1B1, we also generate two other datasets based on existing synthesizers and their human-made presets, and release them to the public with this paper: FM synth timbre\footnote{\url{https://zenodo.org/record/4677102}} 
and subtractive synth pitch.\footnote{\url{https://zenodo.org/record/4677097}}
Using these datasets, we demonstrate new rank-based, synthesizer-motivated evaluation criteria for existing audio representations. Finally, we propose a novel approach to synthesizer hyperparameter optimization, and discuss how perceptually-correlated auditory distance measures could enable new applications in synthesizer design. 

\subsection{Background and Motivation}

\subsubsection{Pre-training and learned representations}

Tobin {\em et al.}~\cite{DBLP:conf/iros/TobinFRSZA17} argue that learning over synthesized data enables transfer learning to the real-world.
Multi-modal training offers additional benefits. 
Multi-model learning of audio---with video in \cite{DBLP:conf/icassp/CramerWSB19} and semantic tags in \cite{drossos:icml:2020}---has led to strong audio representations.
Contrastive audio learning approaches like \cite{saeed2020contrastive} can be used in multi-modal settings, for example by learning the correspondence between a synthesized sound and its underlying parameters.
However, training such models is limited by small corpora and/or the relatively slow synthesis speed of traditional CPU-based synths (Niizumi, p.c.).

\subsubsection{Software Synthesizers}

\begin{sloppypar}
Programming audio synthesizers is challenging and requires a technical understanding of sound design to fully realize their expressive power. Many commercial synthesizers have well over 100 parameters that interact in complex, non-linear ways. One of the most commercially successful audio synthesizers, the Yamaha DX7, is notoriously challenging to program. Allegedly nine out of ten DX7s coming into workshops for servicing still have their factory presets intact \cite{seago2004critical}.
\end{sloppypar}

Since the early 90s, researchers have leveraged advances in ML to develop a deeper understanding of the synthesizer parameter space and to build more intuitive methods for interaction \cite{horner1993machine}. Recently, deep learning has been used for programming synthesizers.  Esling {\em et al.}\ \cite{esling2020flow} trained an auto-encoder network to program the \href{https://u-he.com/products/diva/}{U-He Diva} using 11K synthesized sounds with known preset values. Yee-King {\em et al.} \cite{yee2018automatic} used a recurrent network to automatically find parameters for \href{https://asb2m10.github.io/dexed/}{Dexed}, an open-source software emulation of the DX7.

\subsubsection{Neural Synthesis}

In contrast to traditional synthesis, neural synthesizers generate audio using large-scale machine learning architectures with millions of parameters \cite{engel2017neural}. Differentiable digital signal processing \cite{engel2020ddsp} bridged the gap between traditional DSP synthesizers with the expressiveness of neural networks, exploring a harmonic model-based approach, using a more compact architecture with 100K parameters.
One benefit of synthesized audio is that the underlying factors of variation ({\em i.e.}~the parameters) are known. We combine the ideas of traditional DSP and neural synthesis, yielding a greater level of simplicity and speed by building a GPU-optional modular synthesizer. Our default voice has 78 latent parameters, which model traditional synthesizer parameters. 

\section{Main Contributions}
\label{sec:contributions}

Our synth1B1 corpus and torchsynth software provide a fast, open approach for researchers to do large-scale audio ML pre-training and develop a deeper understanding of the complex relationship between the synthesizer parameter space and resulting audio.
A variety of existing research problems can use synth1B1, including:
\begin{itemize}
\item Inverse synthesis, i.e.\ mapping from audio to underlying synthesis parameters.
\cite{yee2018automatic,esling2020flow}
\item Inferring macro-parameters of synthesizers that are more perceptually relevant. \cite{esling2020flow, tatar2020latent}
\item Audio-to-MIDI. \cite{47659}
\item Perceptual research, such as crafting perceptually motivated auditory representations and inferring timbre dimensions. \cite{vahidi2020timbre}
\end{itemize}
Researchers can also use the synth1B1 corpus to take advantage of innovations from adjacent ML fields, namely: large-scale multi-modal, self-supervised, and/or contrastive learning, and transfer-learning through fine-tuning on the downstream task of interest, particularly tasks with few labeled examples.

\subsection{synth1B1}

synth1B1 is a corpus consisting of one million hours of audio: one billion 4-second synthesized sounds. The corpus is multi-modal in that each sound includes its corresponding synthesis parameters. We use deterministic random number generation to ensure replicability, even of noise oscillators. By default, one tenth of the examples are designated as the test set.

Data augmentation has been used on small-scale corpora to increase the amount of labeled training data. As discussed in \S\ref{sec:intro}, large-scale one-epoch training is preferable, which is possible using synth1B1's million-audio-hours.

Besides sheer size, another benefit of synth1B1 is that it is multi-modal: instances consist of both audio {\em and} the underlying parameters used to generate this audio. The use of traditional synthesis methods allows researchers to explore the complex interaction between synthesizer parameter settings and the resulting audio in a thorough and comprehensive way. 
Large-scale contrastive learning typically requires data augmentation ({\em e.g.}\ image or spectrogram deformations) to construct positive contrastive-pairs \cite{pmlr-v119-chen20j,DBLP:journals/corr/abs-2103-06695}. However, this sort of faux-contrastive-pair creation is not necessary when the underlying latent parameters are known in a corresponding modality.

\subsection{torchsynth}

synth1B1 is generated {\em on the fly} by \href{https://github.com/torchsynth/torchsynth}{torchsynth 1.0}.
torchsynth is an open-source modular synthesizer and is GPU-enabled. torchsynth renders audio at 16200x real-time on a single V100 GPU. Audio rendered on the GPU can be used in downstream GPU learning tasks without the need for expensive CPU-to-GPU move operations, not to mention disk reads. Since it is faster to render synth1B1 {\em in-situ} than to download it, torchsynth includes a replicable script for generating synth1B1.
To accommodate researchers with smaller GPUs, the default batchsize is 128, which requires between 1.9 and 2.4 GB of GPU memory, depending upon the GPU.
If a train/test split is desired, 10\% of the samples are marked as test. Because researchers with larger GPUs seek higher-throughput with batchsize 1024, $9 \cdot 1024$ samples are designated as train, the next 1024 samples as test, {\em etc.} The default sampling rate is 44.1kHz. However, sounds can be rendered at any desired sample rate. %
Detailed instructions are contained in the \href{https://torchsynth.readthedocs.io/}{torchsynth documentation} for the precise protocol for replicably generating synth1B1 and sub-samples thereof. %

\subsection{Questions in Synthesizer Design, and New Pitch and Timbre Datasets and Benchmarks}

To generate a synthesized dataset, one needs to sample the synthesis parameter space. Typically this is achieved through na\"ively sampling parameters uniformly and rendering the resulting audio. Due to the complexity of the parameter space and potential interaction between parameters, such an approach would likely lead to a large number of redundant and/or undesirable sounds (e.g., nearly silent renders, or those having an extreme fundamental frequency). The complexity of this parameter space leads to several open challenges in synthesizer design, specifically focusing on the task of designing and sampling parameters, including:
\begin{itemize}
    \item How do you measure the apparent diversity of a synthesizer's sounds? How do you maximize it?
    \item Is there a parameter sampling strategy that results in audio sounds resembly a human-designed preset?
\end{itemize}
\begin{sloppypar}
The main research barrier to solving these tasks computationally is the lack of an objective auditory distance, i.e.\ a perceptually-relevant audio dissimilarity measure \cite{grey1977multidimensional}. A properly weighted dissimilarity measure could be used, for example, to tune our hyperparameter space to generate sounds that were maximally perceptually different, when torchsynth parameters are randomly sampled.
We devise two auditory-distance evaluation methodologies, and concurrently release two datasets, each representing 22.5 and 3.4 hours of audio respectively, generated by the following open-source synthesizers: a \href{https://github.com/bwhitman/learnfm}{DX7} clone and \href{https://surge-synthesizer.github.io/}{Surge}, as \href{https://zenodo.org/record/4677102}{DOI 10/f7dg} and \href{https://zenodo.org/record/4677097}{DOI 10/f652}, respectively. Importantly, these datasets represent hand-crafted synthesizer sounds---i.e.\ presets designed by humans, not just a computer randomly flipping knobs---which we use in two ways: a) New benchmarks for evaluating audio representations. b) Evaluating the similarity of different sound corpora.
\end{sloppypar}

\section{torchsynth Design}
\label{sec:design-methodology}
\subsection{Synth Modules}
torchsynth's design is inspired by hardware modular synthesizers which contain individual units. Each module has a specific function and parameters, and they can be connected together in various configurations to construct a synthesizer. There are three types of modules in torchsynth: audio modules, control modules, and parameter modules. Audio modules operate at audio sampling rate (default 44.1kHz) and output audio signals. Examples include voltage-controlled oscillators (VCOs) and voltage-controlled amplifiers (VCAs). Control modules output control signals that modulate the parameters of another module. For speed, these modules operate at a reduced control rate (default 441Hz). Examples of control modules include ADSR envelope generators and low frequency oscillators (LFOs). Parameter modules simply output parameters. An example is the monophonic ``keyboard'' module that has no input, and outputs the note midi f0 value and duration.

\begin{figure}

\centering

\begin{subfigure}
    \centering
    \includegraphics[width=0.90\linewidth]{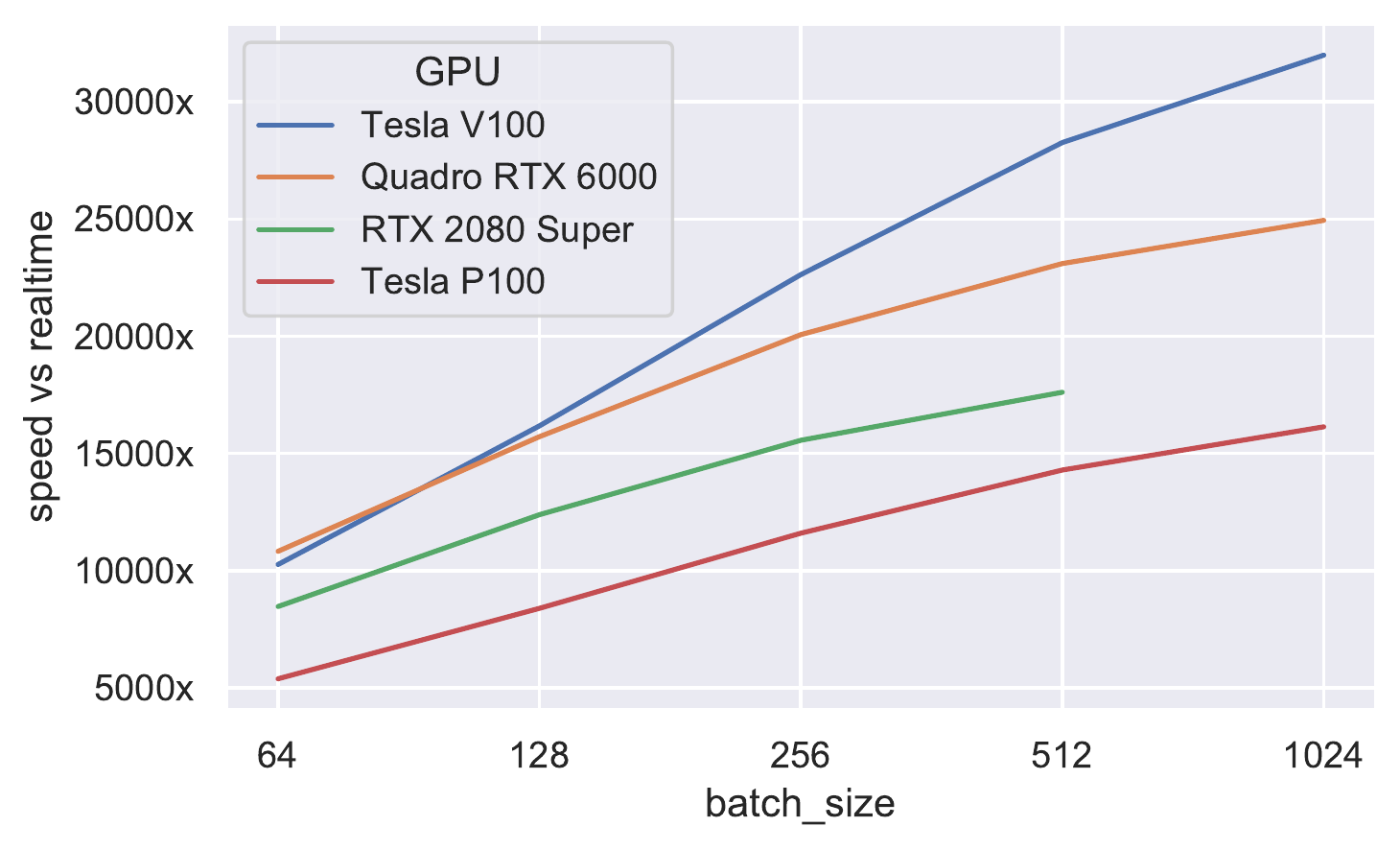}
  \vspace{-1.5em}
    \caption{torchsynth throughput at various batch sizes.}
    \label{fig:gpu-profiles}
\end{subfigure}

\vspace{0.725cm}

\begin{subfigure}
    \centering
    \includegraphics[width=0.85\linewidth]{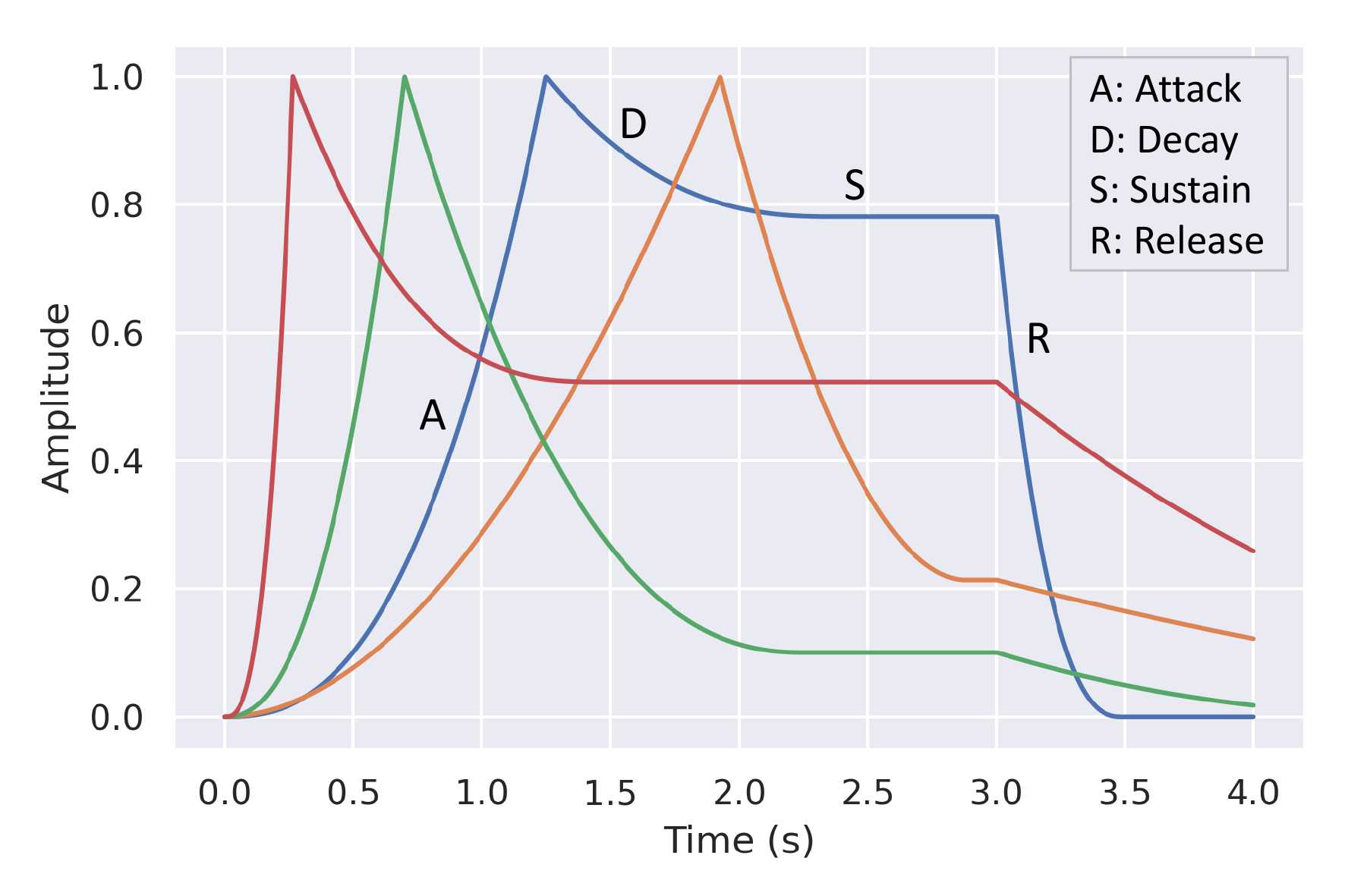}
  \vspace{-1.5em}
    \caption{Batch of four randomly generated ADSR envelopes.
    Each section for one of the envelopes is labelled.}
    \label{fig:adsr}
\end{subfigure}

\vspace{0.725cm}

\begin{subfigure}
    \centering
    \includegraphics[width=0.84\linewidth]{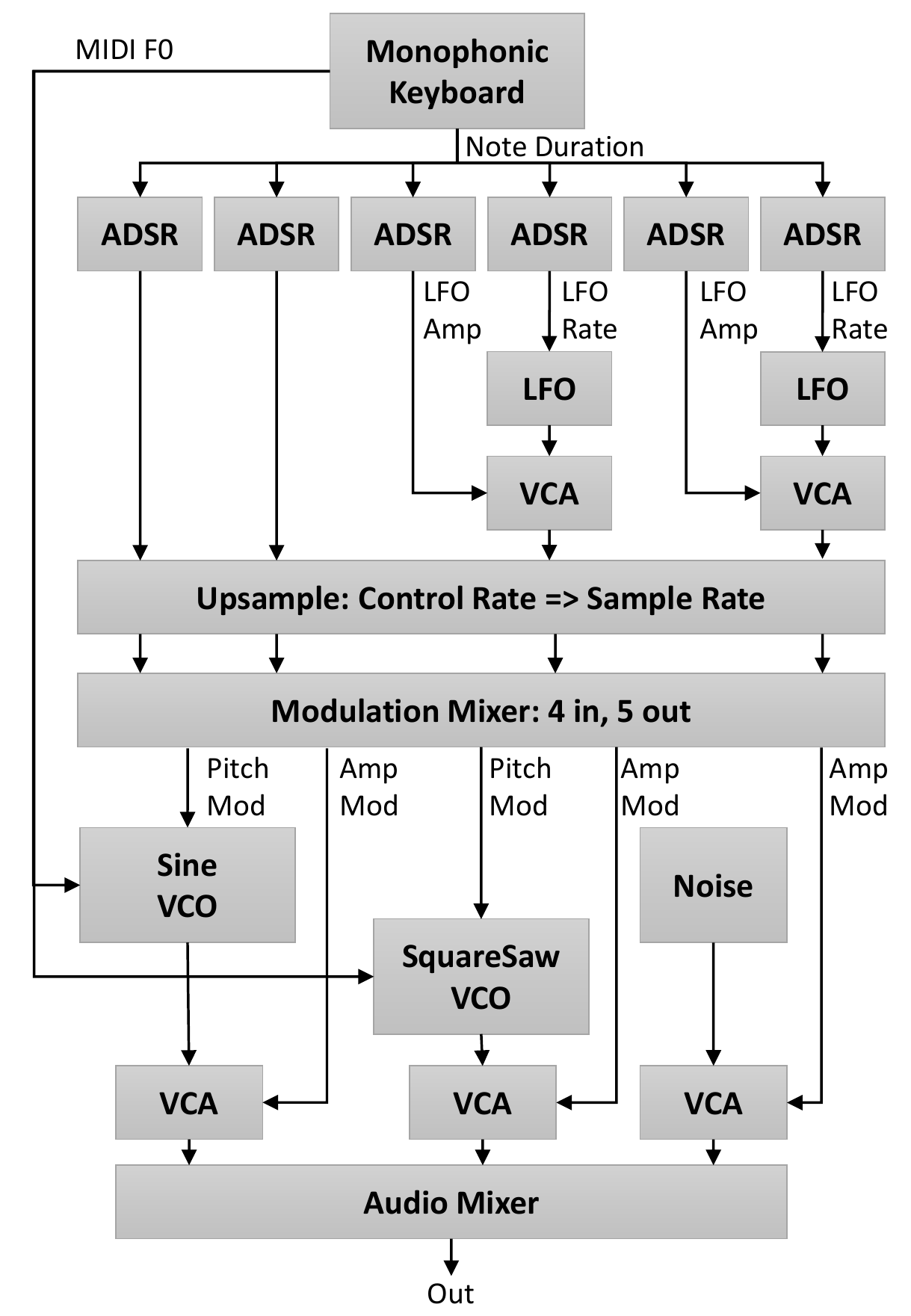}
  \vspace{-1.5em}
    \caption{Module configuration for the Voice in torchsynth}
    \label{fig:voice_diagram}
\end{subfigure}
\end{figure}

To take advantage of the parallel processing power of a GPU, all modules render audio in batches. Larger batches enable higher throughput on GPUs.  Figure~\ref{fig:gpu-profiles} shows torchsynth's throughput at various batch sizes on a single GPU. GPU memory consumption $\approxeq 1216 + (8.19\ \cdot $ batch\_size) MB, including the torchsynth model. The default batch size 128 requires $\approx$2.3GB of GPU memory. 
A batch of 4 of randomly generated ADSR envelopes is shown in Figure~\ref{fig:adsr}.

\subsection{Synth Architectures}

The default configuration in torchsynth is the Voice, which is the architecture used in synth1B1. The Voice is made of the following modules: a Monophonic Keyboard, two LFOs, six ADSR envelopes (each LFO module includes two dedicated ADSRs: one for rate modulation and another for amplitude modulation), one Sine VCO, one SquareSaw VCO, one Noise generator, VCAs, a Modulation Mixer and an Audio Mixer. Modulation signals generated from control modules (ADSR and LFO) are upsampled to the audio sample rate before being passed to audio rate modules. Figure \ref{fig:voice_diagram} shows the configuration and routing of the modules comprised by Voice. %
While the Voice is the default architecture of torchsynth 1.0, any number of synth architectures can be configured using the available modules. A 4-operator frequency modulation (FM) \cite{chowning1973synthesis} synthesizer inspired by \href{https://www.ableton.com/en/packs/operator/}{Ableton Live's Operator instrument} is currently in development.

\begin{figure}[thb]
    \centering
    \includegraphics[width=0.95\linewidth]{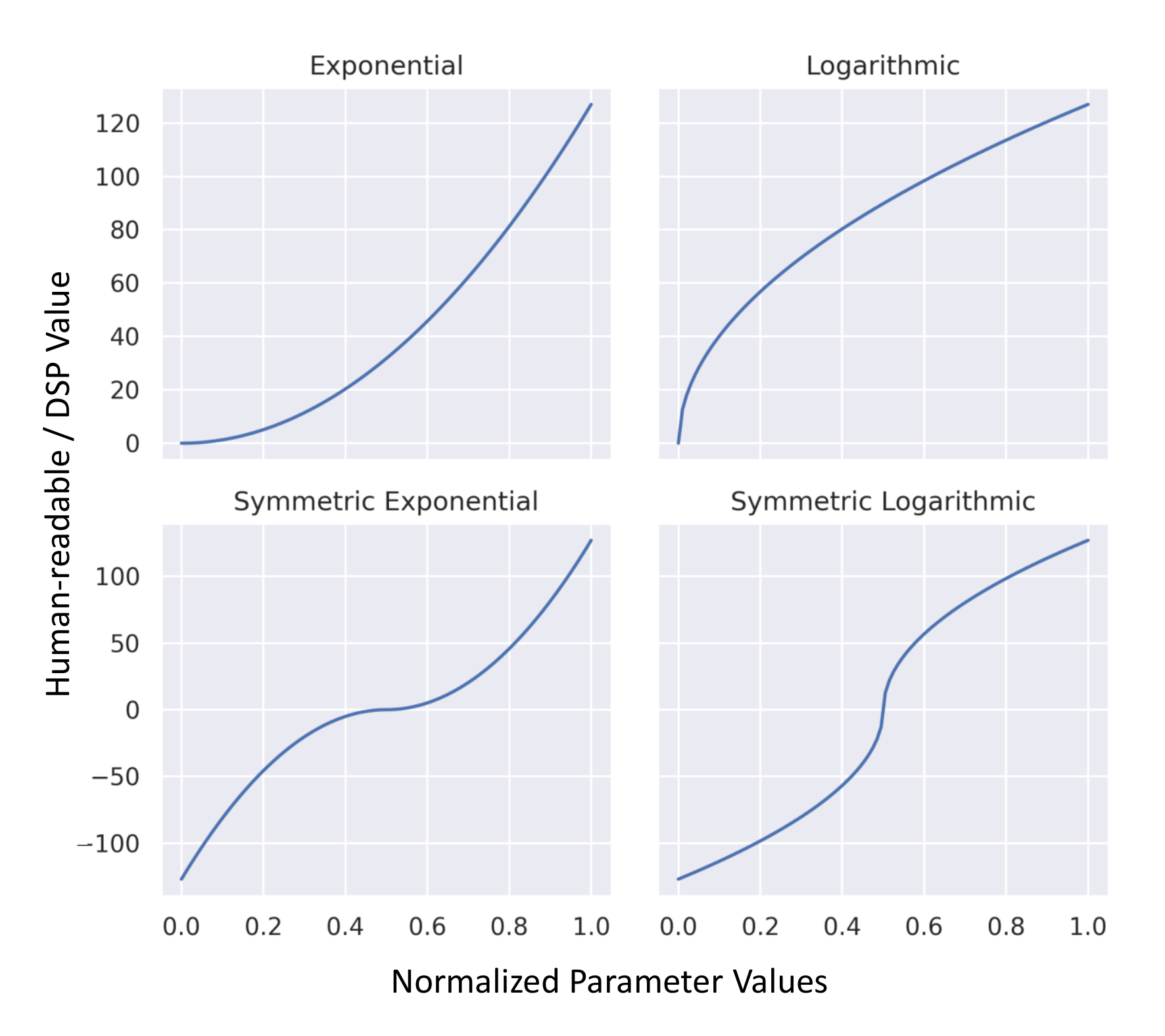}
    \caption{Examples of parameter curves used to convert to and from normalized parameter values and the human-readable values used in the DSP algorithms. The top two curves are non-symmetric curves, mapping to values in the range [0, 127]. The bottom two curves are symmetric, mapping to values in the range [-127, 127].}
    \label{fig:parameter_curves}
\end{figure}

\subsection{Parameters}
\begin{sloppypar}
Module parameters can be expressed in human-readable form with predetermined minimum and maximum values, such as $0 \le \textrm{midi f0} \le 127$. These values are used directly by the DSP algorithms of each module.
Internally, parameters are stored in a corresponding normalized range $[0, 1]$. synth1B1 parameters are sampled uniformly from the normalized range; however, there is potentially a non-linear mapping between the internal range and the human-readable range. Besides the fixed min and max human-readable values, each parameter has two hyperparameters, ``curve,'' and ``symmetry,'' that determine how the internal [0, 1] values are transformed to the human-readable values. The curve can specify a particular logarithmic, linear, or exponential sampling approach. 
Symmetric curves, which alternately emphasize the center or edges of the distribution, are used for parameters that are centered at zero, and take on a range of both positive and negative values (such as oscillator tuning offset). An example set of non-linear curves is shown in Figure \ref{fig:parameter_curves}.
\end{sloppypar}

In our nomenclature, a particular choice of hyperparameter settings, which corresponds to a random sample space of markedly different sonic character, is called a {\em nebula}. The initial Voice nebula was designed by the authors based upon intuition and prior experience with synthesizers. We experiment with tuning the hyperparameters of Voice to generate different nebulae in \S\ref{sec:hyperparameter-tuning}.

\section{Evaluation of Auditory Distances}

We seek (1) to quantify the diversity of sounds that can be generated with torchsynth within a particular nebula; and a similar problem, (2) to quantify to what extent a certain nebula can mimic the variability of sounds in another dataset. In order to do so, we first need a reliable measure of dissimilarity between pairs of sounds, also known as an auditory distance. %

\begin{sloppypar}
Auditory distances typically involve computing some multidimensional representation of a sound, then computing a distance over the representation space. Not all auditory distances are equally informative, depending on what is being measured; the L2 distance between two spectrograms, for example, carries little relative pitch information \cite{turian2020im}. In our case, we are looking to quantify the perceptual diversity of our dataset.
\end{sloppypar}

\begin{table*}[ht]
  \centering
  \begin{tabular}{rlcrrrrrr}
{} & & & \multicolumn{3}{c}{Spearman with a preset} & \multicolumn{3}{c}{DCG across presets} \\
 &  &  &  & Surge & DX7 &  & Surge & DX7 \\
Representation & model choice & normed &  & pitch & velocity & mean & &  \\
\hline
OpenL3 \cite{Cramer:LearnMore:ICASSP:19} & env, mel256, 6144  &              &          \textbf{0.821} &           \textbf{0.746} &              0.896 &             0.880 &              0.908 &                 0.852 \\
OpenL3 \cite{Cramer:LearnMore:ICASSP:19}& env, mel256, 6144 & \checkmark     &          \textbf{0.821} &           \textbf{0.747} &              0.895 &             0.809 &              0.883 &                 0.735 \\
OpenL3 \cite{Cramer:LearnMore:ICASSP:19} & music, mel256, 6144 & \checkmark   &          0.817 &           0.732 &              \textbf{0.903} &             0.820 &              0.916 &                 0.724 \\
OpenL3 \cite{Cramer:LearnMore:ICASSP:19} & music, mel256, 6144  &            &          0.813 &           0.722 &              \textbf{0.903} &             0.892 &              \textbf{0.942} &                 0.842 \\
Coala \cite{drossos:icml:2020} & dual\_ae\_c & \checkmark                    &          0.813 &           0.729 &
 0.896 &             0.555 &              0.547 &                 0.564 \\
Coala \cite{drossos:icml:2020} & dual\_e\_c & \checkmark                     &          0.811 &           0.737 &
 0.884 &             0.569 &              0.576 &                 0.563 \\
 
NSynth Wavenet \cite{engel2017neural} & & \checkmark                            &          0.810 &           0.717 &              \textbf{0.903} &             0.582 &              0.591 &                 0.573 \\
OpenL3 \cite{Cramer:LearnMore:ICASSP:19} & music, linear, 6144 &             &          0.808 &           0.722 &              0.895 &             0.874 &              \textbf{0.943} &                 0.805 \\
OpenL3 \cite{Cramer:LearnMore:ICASSP:19} & music, mel256, 512 &              &          0.804 &           0.710 &              0.899 &             \textbf{0.904} &              \textbf{0.943} &                 \textbf{0.864} \\
OpenL3 \cite{Cramer:LearnMore:ICASSP:19} & music, mel256, 512 & \checkmark    &          0.801 &           0.705 &              0.897 &             0.585 &              0.606 &                 0.564 \\
NSynth Wavenet \cite{engel2017neural} & &                                     &          0.789 &           0.675 &              \textbf{0.903} &             0.835 &              0.893 &                 0.777 \\
Coala \cite{drossos:icml:2020} & dual\_ae\_c &                              &          0.776 &           0.658 &              0.893 &             0.748 &              0.756 &                 0.740 \\
Coala \cite{drossos:icml:2020} & dual\_e\_c &                               &          0.750 &           0.630 &
 0.871 &             0.681 &              0.710 &                 0.652 \\
Multi-scale spectrogram \cite{engel2020ddsp,steinmetz2020auraloss} & linear+log, [4096 ... 64] & &          0.792 &           0.690 &              0.894 &             0.543 &              0.555 &                 0.531 \\
Multi-scale spectrogram \cite{engel2020ddsp,steinmetz2020auraloss} & log, [4096 ... 64] & &          0.786 &           0.689 &              0.884 &             0.542 &              0.566 &                 0.518 \\
Multi-scale spectrogram \cite{engel2020ddsp,steinmetz2020auraloss} & linear, [4096 ... 64] & &          0.658 &           0.410 &              \textbf{0.905} &             0.447 &              0.343 &                 0.551 \\
Coala \cite{drossos:icml:2020} & cnn & \checkmark                          &          0.555 &           0.303 &              0.806 &             0.485 &              0.433 &                 0.537 \\
Coala \cite{drossos:icml:2020} & cnn &                                   &          0.552 &           0.297 &              0.807 &             0.714 &              0.614 &                 0.815 \\
 \end{tabular}
  \caption{Performance of representations on experiments defined in \S~\ref{sec:experiment1} and \ref{sec:experiment2}. Best scores, and scores within 0.002 of the best, are bold-faced. $\ell_1$ distance was used because it outperformed $\ell_2$. We sort by mean spearman within a preset.
  }
  \label{tab:distance-eval}
\end{table*}

\subsection{Additional Datasets}
To find a suitable distance, we devised two experiments using two new datasets.
Sounds in each of the following datasets are RMS-level normalized using the \href{https://github.com/kklobe/normalize}{normalize} package.

\subsubsection{DX7 Timbre Dataset}
\label{sec:dx7}

Given 31K human-designed presets for the DX7, we generated 4-second samples on a fixed midi pitch (69 = A440) with a note-on duration of 3 seconds (using \href{https://github.com/bwhitman/learnfm}{this DX7 clone}). For each preset, we varied only the velocity, from 1--127. This dataset is built on the assumption that velocity
effects a meaningful, monotonic variation in timbre when it is explicitly programmed into a DX7 patch. Not all DX7 patches are velocity sensitive, and some are more sensitive than others. In our generation process, sounds that were completely identical---{\em i.e.} each sample matched with error 0---were removed. 8K presets had only one unique sound. The median was 51 unique sound per preset, mean 41.9, stddev 27.4.

\subsubsection{Surge Pitch Dataset}
\label{sec:surge}

To explore a second dimension of variability, in this case {\em pitch}, we used the \href{https://github.com/surge-synthesizer/surge-python}{Surge synthesizer Python API} and the 2.1K standard Surge presets. The open-source \href{https://surge-synthesizer.github.io/}{Surge synthesizer} is a versatile subtractive synthesizer with a variety of oscillator algorithms. 
Here we held the velocity constant at a value of 64, and varied midi pitch values from 21--108, the range of a grand piano. Only a small percentage of presets (like drums and sound effects) had no meaningful pitch variation, and thus no perceptual ordering as pitch increases.
Therefore, a small fraction of presets are unclassifiable, imposing a uniform upper bound in accuracy across the board for all auditory distances.

\subsection{Experiments}
\subsubsection{Distance Experiment 1: Timbral and Pitch Ordering Within a Preset}
\label{sec:experiment1}

In this experiment, we measure the ability of an auditory distance to order sounds by ``timbre,'' or by pitch, in the DX7 and Surge datasets, respectively. In effect, the experiment is two evaluations in parallel, run on two separate datasets.

We sample a random preset with at least 3 unique sounds. For each sound $s$, we pick a random sound $s_l$ from this preset with a lower rank (using the DX7 set, this would be a sound having the same pitch but a lower velocity; for the Surge dataset this is a sound having the same velocity but lower pitch); and a random sound $s_h$ with higher rank.

For each of $s$, $s_l$ and $s_h$, we compute the distance $d(\cdot, \hat{s})$ between this sound and all other sounds $\hat{s}$ in the dataset. While $s$ is the sample of interest, distance measures are strictly non-negative. Therefore, we seek a concurrent metric to determine whether the compared sound $\hat{s}$ is ``above'' or ``below'' $s$. If the sound $\hat{s}$ is closer to $s_l$, we determine the sign of the distance to $s$ to be negative. If $\hat{s}$ is closer to $s_h$, we determine the sign of the distance to $s$ be positive. As a result, we have a signed distance metric comparing the sound $s$ to every other sound in the dataset.

This set of distances is then correlated to the ground-truth index of pitch, or velocity (depending on the dataset). The correlation, here a Spearman rank correlation, reflects the extent to which the signed distance can properly order the dataset by variability in pitch or velocity. 
One limitation of this methodology for inducing a forced ranking from simple distance is that if, say, $s = 80, s_l = 31, s_h = 81,$ and $\hat{s} = 79$, we might judge $\hat{s}$ as closer to $s_h$ and thus above $s$. We controlled for this by using the same choice for every auditory distance of $s_h$ and $s_l$ given $s$.

Formally, we estimate:
\begin{align}
\begin{split}
\mathop{\mathbb{E}}_{S \in P, s \in S, s_l, s_h \sim S, s_l < s < s_h} \bigg [ \mathop{\rho}_{\hat{s} \in S}\Big (& \textrm{rank}(\hat{s}), d(s, \hat{s}) \ \cdot \\
& \textrm{sgn} \big (d(s_h, \hat{s}) < d(s_l, \hat{s} \big ) \Big) \bigg ]
\end{split}
\label{eq:spearman}
\end{align}
$P$ is the set of presets, $S$ sounds in that preset, and $\rho$ is spearman.

\subsubsection{Distance Experiment 2: Determine a Synthesis Preset}
\label{sec:experiment2}

In this constrained environment, a good distance measure should have a relatively low distance between sounds generated by the same preset. For each trial, we sample 200 different presets. We sample 2 unique sounds from each preset. For each sound, we compute its distance against the 399 other sounds, and then compute the discounted cumulative gain (DCG) \cite{pmlr-v30-Wang13} of the sound from the same preset, with binary relevance. The DCG is computed for all 400 sounds in the trial. We perform 600 trials.

In the pitch dataset, to control for the helical nature of pitch perception \cite{shepard1982geometrical}, the second sound was always an interval of six semitones (AKA a tritone, {\em diabolus in musica}) 
from the first note. This ensures that pitches avoid similar partials due to overlapping harmonics that could be easily matched.

\subsection{Evaluation Results}

We audition the following distances: a) The multi-scale spectrogram distance has been used in a variety of applications, particularly in speech synthesis \cite{wang_neural_2019,DBLP:conf/icassp/YamamotoSK20} but also in music \cite{dhariwal2020jukebox,engel2020ddsp,DBLP:journals/corr/abs-2008-01393};
b) NSynth Wavenet \cite{engel2017neural} is a Wavenet-architecture trained on NSynth musical notes; c) OpenL3 \cite{Cramer:LearnMore:ICASSP:19} was  trained multi-modally on AudioSet audio and video, on two distinct subsets: music and environmental sounds; d) Coala \cite{drossos:icml:2020} was trained multi-modally on Freesound audio and their corresponding tags.

We experimented with a variety of hyperparameter settings for the representations. The best results are in Table~\ref{tab:distance-eval}. We use $\ell_1$ distance because it gave better results than $\ell_2$ across the board. For Coala and NSynth Wavenet, normalizing
improves the spearman scores, but harms the DCG across presets. Normalization had little effect on OpenL3. %
OpenL3 (music, mel256, 512) achieves the best score on DCG across presets, and its compactness makes it an appealing choice for the remaining experiments in the paper.

\section{Similarity between Audio Datasets}
\label{sec:similarity-dataset}

To evaluate the similarity between two datasets of audio samples $X$ and $Y$, we use the maximum mean discrepancy (MMD) \cite{JMLR:v13:gretton12a}. We use the following MMD formulation, assuming $X$ and $Y$ both have $n$ elements:
\begin{equation}
    \textrm{MMD}(X, Y) = \frac{1}{n n} \sum_{i,j=0}^n 2 \cdot d(x_i, y_j) - d(x_i, x_j) - d(y_i, y_j)
  \label{eq:mmd}
\end{equation}
MMD allows us to use our chosen distance measure---OpenL3 (music, mel256, 512) $\ell_1$---as the core distance $d$.

\begin{table}[th]
    \centering
    \begin{tabular}{r|r|ll}
MMD & std & corpus 1 & corpus 2 \\
\hline
4.396 & 0.123 & white & white \\
21.409 & 4.729 & dx7 & dx7 \\
23.732 & 3.615 & FSD50K & FSD50K \\
24.130 & 5.251 & torchsynth & torchsynth \\
27.824 & 9.821 & surge & surge \\
2751.519 & 80.955 & torchsynth & surge \\
2884.843 & 67.264 & surge & dx7 \\
3001.857 & 71.888 & torchsynth & FSD50K \\
3637.845 & 79.265 & torchsynth & dx7 \\
4756.952 & 112.705 & surge & FSD50K \\
7413.105 & 111.897 & dx7 & FSD50K \\
13202.202 & 61.558 & white & FSD50K \\
16985.319 & 92.992 & white & torchsynth \\
18488.926 & 67.277 & white & surge \\
20374.929 & 78.886 & white & dx7 \\
\end{tabular}
    \caption{MMD results comparing different audio sets, including the stddev of the MMD over the 1000 trials.}
    \label{tbl:corpora-mmd}
\end{table}

For Surge and DX7, we selected sounds with midi pitch 69 and velocity 64. We also generated a set of 4-second samples of white-noise, and used excerpts from the FSD50K evaluation set \cite{fonseca2020fsd50k}, which is broad-domain audio, trimmed to 4 seconds. From each corpus, we randomly sampled 2000 sounds, to match the size of the smallest corpus (Surge). We performed 1000 MMD trials, each time comparing $n=1000$ sounds from one corpus to $n=1000$ sounds from another,  randomly sampled each trial. To estimate the diversity within a particular corpus, we evaluated MMD over 1000 distinct 50/50 partitions of the corpus.

Table~\ref{tbl:corpora-mmd} shows the result of average MMD computations between different audio corpora. 0.0 would be perfectly identical, only occurring if the two corpora had identical sounds. Some results are expected, whereas some are counter-intuitive and suggest potential issues in our use of the OpenL3 distance measure. These results are sometimes perceptually incoherent, and suggest that the use of the auditory distance measures explored may impede progress in automatic synthesizer design, as we will illustrate in the following section.

\begin{itemize}
    \item White-noise is the most similar to itself of all comparisons.
    \item FSD50K broad-domain sounds are, strangely, considered to have less within-corpus diversity than torchsynth or Surge sounds. However, the variance is high enough that it is hard to have statistical confidence in this unexpected result.
    \item More troubling are low-variance estimates that torchsynth is more similar to FSD50k than it is to the dx7 synth. {\em A priori}, one would expect that synthesizers would sound more similar to each other than broad domain audio. %
    \item White noise is the least similar to DX7 synth sounds of all corpora.
\end{itemize}

\section{torchsynth Hyper-Parameter Tuning}
\label{sec:hyperparameter-tuning}

How can we guarantee the maximum diversity of sounds within a particular nebula? Similarly, to what extent can torchsynth adopt the characteristics of a separate given corpus of audio? Recall from \S\ref{sec:design-methodology} and Figure~\ref{fig:parameter_curves} that for each module parameter, the choice of scaling curve is a hyperparameter. Initial hyperparameters were chosen perceptually and based upon prior-knowledge of typical synth design.

In principle, we can use MMD (Equation~\ref{eq:mmd}) as an optimization criterion to tune these hyperparameters a) to maximize sonic diversity; or b) model the characteristics of another dataset. We use Optuna \cite{optuna_2019}, initializing with 200 random grid-search trials, and subsequently using CMA-ES sampling for 800 trials. In each trial, we generate 256 random torchsynth sounds with the Optuna-chosen hyperparameters. %
Hyperameter curves were sampled log-uniform in the range [0.1, 10]. The top 25 candidates were re-evaluated using 30 different MMD trials, to pick the best hyperparameters. However, MMD estimates are only as good as the underlying similarity metric (OpenL3-$\ell_1$) that it uses.

For these experiments, the authors and non-author musicians conducted blinded listening experiments of the tuned nebula and our manually-chosen nebula, and listened to 64 random sounds. Only after independent qualitative evaluation did we unblind the nebulae.

\subsection{Restricting hyperparameters}

Many torchsynth 1.0 Voice sounds (the default nebula) have an eerie sci-fi feel to them.
To find the drum nebula, we used Optuna to choose hyperparameters to {\em minimize} the OpenL3-$\ell_1$-MMD against 10K one-shot percussive sounds \cite{ramires2020}. In this experiment, no hyperparameters were frozen; all were permitted to be tuned at once. 
We had hoped to find that OpenL3-$\ell_1$-MMD would find appropriate percussive curves.

Overall, we found the sounds of the resulting drum nebula unpleasant to listen to.
Many of the sounds did not resemble percussion, and others made use of extreme use of high and low oscillator tunings. %
We suspect that the many wide pitch-modulation sweeps present in the resulting audio were an attempt on the part of the optimizer to match the broadband energy in the target drum transients.

Curious to see if this negative result was due to a failure of the distance measure, or instead a systemic limitation in the design of the torchsynth 1.0 Voice, we next hand-tuned the hyperparameters to create a sensible drum nebula, which is shared as part of our repository. Many of the resulting sounds have a quality akin to early drum machines, and the distribution of sounds is overall much more percussion-like. We encourage the reader to listen to this nebula, which is available on \href{https://torchsynth.readthedocs.io/}{torchsynth site}.

The process of hand-designing this drum nebula revealed to the authors one clear limitation in torchsynth 1.0: all synth parameters are sampled independently. In kick drums, for example, low-tuned oscillators tend to correlate to short, snappy envelope settings; but such envelopes are not appropriate for all percussive sounds. In future work, we are interested in investigating multivariate sampling techniques, which would allow more focused cross-parameter modal sound sampling.

\subsection{Maximizing torchsynth diversity}
\label{sec:diversity}

We also attempted to tune our hyperparameters to maximize torchsynth MMD, i.e.\ increase the perceptual diversity of sounds generated by torchsynth itself. As before, Optuna was used to choose hyperparameters that maximized the OpenL3-$\ell_1$-MMD and thus increase the diversity of sounds. Nonetheless, the ``optimized'' nebula exhibited pathologies in pitch, favoring extremely low and high pitches. It may be that OpenL3-$\ell_1$ overestimates perceptually diversity in these frequency ranges. We performed numerous experiments restricting the hyperparameters Optuna was permitted to modify, such as prohibiting changes to midi f0 and VCO tuning and mod depth. Consistently, listeners preferred our manually design nebula to automatically designed ones in blind tests. We consider this another important negative result that points to the need for further work in automatic synthesizer design.  %

\section{Open Questions, Issues, and Future Work}
\label{sec:issues}

Many of our experiments in automatic synthesizer design hinge on having a perceptually-relevant auditory distance measure. OpenL3 (music, mel256, 512) $\ell_1$ performed well on our quantitative synthesizer experiments (Table~\ref{tab:distance-eval}), but exhibited some issues in the context of this task in qualitative listening tests, in particular its insensitivity to extreme pitch and inability to model percussion.

Learning a perceptually-relevant auditory distance measure is an open research question. Manocha {\em et al.}\ \cite{Manocha:2020:ADP} use manually-annotated ``just noticeable differences'' (JND) trials generated using active learning to induce a perceptual distance measure. However, these experiments only work with speech and do not include pitch variations, so their model was inappropriate for our task. 

The lack of a perceptually accurate auditory distance measure, at least in the context of this task, prevented us from precisely estimating the perceptual diversity of sounds that is expressible by torchsynth, as well as other synthesizers like Surge and DX7. 
We present, then, the following open question: How do we craft an auditory distance measure that can {\em perceptually} measure (and thereby optimize) synthesizer diversity, or similarity to an existing sound corpus?

A perceptually-relevant auditory distance measure for music opens the door to many possible advances in synthesizer design, including: estimating and maximizing the diversity of synthesizer, mimicking existing synthesizers through automation, inverse synthesis, automatic transcription, and the other tasks described in \S\ref{sec:contributions}.

torchsynth 1.0 focuses on high throughput and creating a diverse synth1B1 dataset. %
There are a handful of improvements we want to add to torchsynth: 1) Stress-tested differentiable modules, 2) Subtractive filters, 3) Additional architectures including FM synthesis, 4) Multivariate parameter selection, 5) High-throughput modules that resemble human speech, and 6) A standardized modular architecture for high-throughput audio {\em effect} research.

Despite these open questions, we believe that the synth1B1 corpus is a significant and useful contribution to the world of audio ML research, for its enormous size, speed, and corresponding multi-modal latent parameters.

\section{Conclusions}

We release synth1B1, a multi-modal corpus of synthesizer sounds with their corresponding latent parameters, generated on-the-fly 16200x faster than realtime on a single V100 GPU. This corpus is 100x bigger than any audio corpus present in the literature. Accompanying this dataset is the open-source modular GPU-optional torchsynth package. We hope that larger-scale multi-modal training will help audio ML accrete the benefits demonstrated by previous NLP and vision breakthroughs.

We freely release pitch and timbre datasets based upon human-designed synthesizer presets, and novel evaluation tasks on which we benchmark a handful of audio representations. We present several novel research questions, including how to estimate and maximize the diversity of a synthesizer, as well as how to mimic existing synthesizers. We outline issues and open research questions that currently impede this sort of experimental work, in particular demonstrating negative results of auditory distance measures.

\section{Acknowledgments}

\begin{sloppypar}
The authors wish to thank Alexandre Défossez, Carson Gant, Colin Malloy, Dan Godlovitch, Gyanendra Lucky Das, Humair Raj Khan, Nicolas Pinto, and Will Maclellan and the reviewers for their help.
\end{sloppypar}

\nocite{*}
\bibliographystyle{IEEEbib}
\bibliography{ddsp}

\end{document}